\documentclass[aps,prb,twocolumn]{revtex4}
\usepackage{graphicx}

\begin{document}
\title{ $\rm Na_2V_3O_7$, a frustrated nanotubular system with
  spin-1/2 diamond rings} 

\author{T. Saha-Dasgupta$^1$, Roser Valent{\'\i}$^2$, F. Capraro$^2$, C. Gros$^2$}
\affiliation{$^1$
S.N. Bose National Centre for Basic Sciences, JD
Block, Sector 3, Salt Lake City, Kolkata 700098, India}
\affiliation{$^2$
Institut f{\"u}r Theoretische Physik, Universit{\"a}t Frankfurt,
D-60438 Frankfurt, Germany}
%\affiliation{$^3$
%Department of Physics, University of the Saarland,
%66041 Saarbr{\"u}cken, Germany}
%\pacs{75.30.Gw, 75.10.Jm, 78.30.-j }
\date{\today}

\begin{abstract}
Following the recent discussion on the puzzling nature of the
interactions in the nanotubular system Na$_2$V$_3$O$_7$, we
present a  detailed {\it ab-initio} microscopic analysis of its
electronic and magnetic properties.  By means of a non-trivial
downfolding study we
propose an
effective model in terms of tubes of  
nine-site rings with the geometry of a spin-diamond necklace
with frustrated inter-ring interactions.
We show that this model  provides a quantitative account of the
observed magnetic behavior.
\end{abstract}
\maketitle
%%%%%%%%%%%%%%%%%%%%%%%%%%%%%%%%%%%%%%%%%%%%%%%%%%%%%%%%%%%%%%%%%%%
%%%%%%%%%%%%%%%%%%%%%%%%%%%%%%%%%%%%%%%%%%%%%%%%%%%%%%%%%%%%%%%%%%%
{\it Introduction-}
Low-dimensional quantum spin systems with
chain, ladder or planar geometries have attracted much attention in
the last years \cite{review_lemmens} due to their  unconventional
magnetic properties.
Recently substantial effort has been devoted to the search for new materials
with exotic topologies and novel properties.  For instance,  the  group of
Millet {\it et al.} \cite{millet} has succeeded in synthesizing
the first transition-metal-oxide-based nanotubular system  
Na$_2$V$_3$O$_7$. 
%%This compound is a member of the quarter-filled
%%ladder compound NaV$_2$O$_5$ \cite{review_lemmens}, which has
%%been intensively studied due to its unique coupling between
%%spin and pseudospin degrees of freedoms. 
The discovery of Na$_2$V$_3$O$_7$ has introduced a new class of novel
geometric structures in the field of magnetic nanostructures.

\begin{figure}
\includegraphics[width=8.5cm,keepaspectratio]{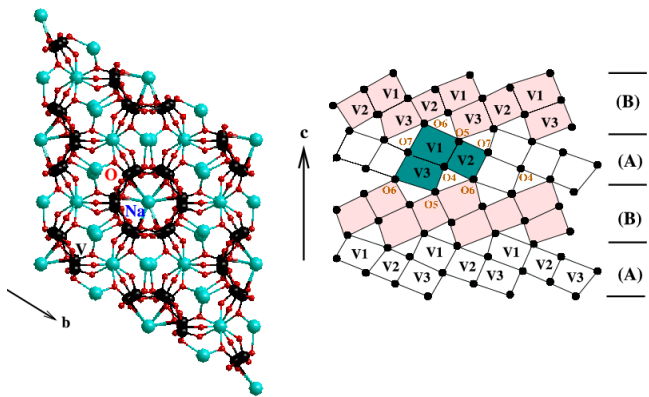}
\caption{(Color on-line) Crystal structure of Na$_{2}$V$_{3}$O$_{7}$. (a) Projection
on to (001) plane. V, Na and O atoms are denoted by black, blue and red
balls
respectively. (b) Unfolded view of the edge and corner-sharing
VO$_{5}$ pyramids within a tube.} 
%%The basic V1-V2-V3 unit formed on the basis of short first
%%nearest neighbor V-V distances are marked in dark green.}
\label{structure}
\end{figure}

The V$^{4+}$O$_{5}$ pyramids
% (V's are in nominal valence state 4+)
in
Na$_2$V$_3$O$_7$ share edges and corners to form a
nanotubular structure - a geometry first discovered for carbon-
with Na atoms  located inside and around each
individual tube (see FIG.\ \ref{structure}).
The complex geometry of
the compound  is expected to provide  non-trivial paths for the
exchange interaction and hence to exhibit a non-trivial magnetic behavior.
Initially  this compound was predicted to be an example of
a
$S=1/2$ nine-leg (or three-leg ladder) system with periodic boundary conditions
along the rung direction\cite{millet}.
  Using the extended H{\"u}ckel method, Whangbo and Koo \cite{Whangbo00}
conjectured, on the other hand, that  the tubes could
be described by six mutually intersecting
helical spin chains which should show a gap in the spin excitation spectra.

Unlike antiferromagnetic (AF) spin ladders with even number of legs,
the behavior of odd-leg ladders is strongly influenced by
topology.
 An odd-leg ladder with open
boundary conditions in the rung direction behaves effectively as a $S=1/2$ AF
Heisenberg chain and therefore shows no gap in the spin-excitation spectra.
This situation changes dramatically when the boundary condition along
the rung direction is  periodic, as is the case for the spin tubes in
Na$_2$V$_3$O$_7$. The introduction of periodic boundary conditions
brings in an additional degree of freedom in the problem, namely the
chirality\cite{Subrahmanyam94,Schulz_96,Kawano97,Wang01} and the ground state is then
four-fold degenerate with a gap to the first  excited state.
However,  the magnetic properties of Na$_2$V$_3$O$_7$
as reported in the literature neither fit
the properties of odd-leg spin tubes nor to that of
helical spin chains as proposed
by Whangbo and Koo \cite{Whangbo00} since no appreciable spin gap
could be detected in the susceptibility measurements\cite{Gavilano03}.
%% {\it
%%Na$_2$V$_3$O$_7$  exhibits  neither a phase transition to
%%a magnetically ordered state nor spin-freezing phenomena at
%%temperatures above 2 K.}
A way out of this puzzling situation has been suggested by L\"{u}scher
{\it et. al.}\cite{Luscher04} by considering frustrated inter-ring
couplings within the framework of a three-leg spin tube. While this surely
provides an interesting  model, its relevance
for Na$_2$V$_3$O$_7$  remains yet to be examined.

Eventhough this compound has generated interest and controversy,
no first-principles microscopic study has been reported so far. Our
work in that respect is the first microscopic
study carried out for  Na$_2$V$_3$O$_7$.  We have
analyzed the {\it ab-initio} density functional theory (DFT) results in terms
of the newly developed N-th order muffin-tin-orbital (NMTO) based downfolding
technique\cite{nmto} to provide a microscopically-derived spin model for Na$_2$V$_3$O$_7$.
Our results show that the appropriate description
of the system is that of
tubes of  
nine-site rings
with partially frustrated, next-nearest neighbor intra-ring
interactions and  frustrated  inter-ring interactions of weaker strength.
Validity of this model is provided by the good agreement
of our calculated susceptibility data with the measured data.

{\it Ab-initio Study-}
Na$_2$V$_3$O$_7$ crystallizes\cite{millet}
in the trigonal space group, P31${c}$.  The
unit cell contains six formula units
and the lattice constants are  $a=$ 10.89 {\AA}\  and
$c=$ 9.54 {\AA}. There are three inequivalent V sites -V1,
V2, V3, seven inequivalent O sites - O1-O7 and four inequivalent
Na sites, Na1- Na4.  Each V is surrounded by five oxygen atoms
with one characteristic short V-O bond
giving
rise to distorted square pyramids which
are connected to each other to form nanotubes with internal
diameter of about 5 {\AA}   (see  
FIG.\ \ref{structure}(a)).
The arrangement of VO$_{5}$ pyramids in a single tube is better viewed in
the unfolded idealized representation of the nanotube as shown in
FIG.\ \ref{structure}(b).
The O1, O2 and O3 atoms, not marked in the figure,
are positioned at the apex of VO$_{5}$ pyramids centered by V1, V2 and V3 atoms
respectively, all of them pointing out of the tube.
The edge sharing VO$_{5}$ pyramids form nine-member rings
out of the basic unit V1-V2-V3, in alternating sequence
of (V2-V1-V3)-(V2-V1-V3)-(V2-V1-V3) and (V3-V1-V2)-(V3-V1-V2)-(V3-V1-V2)
within slices (A) and (B). Rings in
slices (A) and (B) are connected to each other by the corner sharing oxygens
O5 and O6 to form a nanotube oriented along the crystallographic c-direction.
%The Na atoms provide cohesion of the network.
%===========================================================================
%===========================================================================

The DFT band-structure in the  
generalized gradient approximation (GGA)\cite{Perdew_96},
calculated using the linearized muffin tin orbital (LMTO) method
based on the Stuttgart TBLMTO-47 code\cite{Andersen_75}
and the linear augmented plane wave (LAPW)\cite{WIEN2K}
method, consists of primarily oxygen$-p$  and V$-d$
derived bands in the region of interest\cite{footnote}.
The oxygen$-p$ dominated bands are separated by an energy gap
of about 3 eV from the  V$-d$ dominated
bands . The V$-d$ dominated
bands span an energy range of about 4 eV, starting from about -0.7 eV to
3.5 eV, with the zero of energy set at the GGA Fermi energy.
The pyramidal co-ordination of oxygen atoms surrounding the V atom, sets the V$-xy$
$d$ orbital to be the lowest energy state within
the V$-d$ manifold   (in the local reference frame, with the z-axis
pointing along the  shortest V-O bond and the x-axis pointing along
the next shorter V-O bond). The V$-xy$ bands
therefore appear as {\it essentially} non-bonding set of bands  extending
from  about $-0.7\,{\rm eV}$ to $0.4\,{\rm eV}$ around
the Fermi level, with small mixing of oxygen character.
The crystal-field split V$-yz$, $xz$,
$3z^{2}-1$ and $x^{2}-y^{2}$ complexes appear in the energy spectrum above
the V$-xy$ complex in order of increasing energy.
Na-derived states lie farther high up in energy with practically no
mixing to bands close to the Fermi energy \cite{footnote1}.

%===========================================================================
%\begin{figure}
%\vskip 2in
%\includegraphics[width=5cm]{navo_bands.eps}
%\vspace{-0.3cm}
%\caption{Energy bands for Na$_{2}$V$_{3}$O$_{7}$ obtained
%from DFT. The bands are plotted along the high symmetry directions with
%$\Gamma = (0,0,0)$, $K = (-1/3,2/3,0)$, $M = (0,1/2,0)$, $A = (0,0,1/2)$,
%$L = (0,1/2,1,2)$
%and $H = (-1/3,2/3,1/2)$.}
%\label{bands}
%\end{figure}
%===========================================================================
Starting from such a density-functional input, it is a non-trivial task
to build up a low-energy  model Hamiltonian relevant for the system. For the sake of
uniqueness, however,  it is essential for such model Hamiltonians to be derived in
a first principles manner containing the essential chemistry of the material. In recent
years the N-th-order-muffin-tin-orbital-method-based\cite{nmto}
downfolding technique has
been
successful in achieving this goal\cite{nmto_app}.  
The method relies on designing energy-selective
Wannier-like effective orbitals by integrating out degrees of freedom
that are not relevant  - a method called {\it downfolding}.
The
few-orbital Hamiltonian is then constructed
in the basis of these Wannier-like effective orbitals. Since the degrees
of freedom are integrated out, rather than being simply ignored, the method
naturally takes into account
the renormalization effect coming from orbitals not considered explicitly.
In particular, in the present case, we integrate out all the degrees
of freedom other than V$-xy$ orbitals.
The effective V$-xy$ muffin tin orbitals (MTO's) generated
in the process contains in its tail the integrated out O$-p$ and remaining V$-d$
orbitals, the weight being proportional to their mixing to V$-xy$ derived
bands. Fourier transform
in the
{\it downfolded} V$-xy$ basis gives the tight-binding Hamiltonian,
$
H_{TB}=-\sum_{<i,j>}t_{ij}(c^{\dagger}_jc_i+c^{\dagger}_ic_j)
$
in
terms of dominant V-V effective hopping integrals, $t_{ij}$,
where $i$ and $j$ denote a pair of V$^{4+}$ ions.  

The nearest neighbor (n.n.) V-V interactions within the V1-V2-V3 basic
unit
(marked by dark green in FIG.\ \ref{structure}(b)) proceed via the edge-sharing oxygens while
the  intra-ring  next nearest neighbor (n.n.n.) interactions proceed
via corner-sharing oxygens \cite{note}.
Our first-principles-derived hopping integrals
show that the edge-sharing n.n. V-V interactions
within the V1-V2-V3 basic unit, $t_1$, $t_2$ and $t_3$ (see FIG. 2 and TABLE \ref{table}), are of magnitude ranging from
-0.14 eV to -0.18 eV.  The n.n.n corner sharing intra-ring
interactions $t'_1$ and  $t'_2$   are nearly equally strong.
The inter-ring V-V
couplings $t_{i\perp}$,
$t_{i\perp}'$  $i=1,2$ are an order of magnitude weaker than the intra-ring
couplings.
%% with their magnitudes ranging from -0.03 to -0.02 eV.
Nevertheless,  we notice that the n.n.n inter-ring interactions ($t^{'}_{1\perp}$ and
$t^{'}_{2\perp}$)  are equally strong
as the n.n.  inter-ring interactions ($t_{1\perp}$ and
$t_{2\perp}$). This induces inter-ring frustration
which, as we will  see later, could be of fundamental importance
for the description of the   magnetic
behavior of the system at low temperatures.
%===========================================================================
\begin{figure}
\includegraphics[width=6.5cm,keepaspectratio]{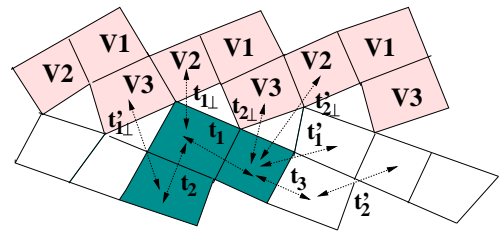}
\caption{Hopping parameters for  Na$_{2}$V$_3$O$_7$ for V atoms
  belonging to slices (A) and (B) in FIG.1(b). For identification of
  V atoms belonging to slice (A) refer to FIG.1(b).
}
\label{hopping}
\end{figure}

%===========================================================================
\begin{table}
\caption{Inter and intra-ring hopping integrals in eV.
% Only hoppings of
%magnitude larger than 0.01\,eV are shown. 
The next important hoppings
are about 0.001 eV.\label{table}}
\begin{ruledtabular}
\begin{tabular}{cccccccc|cccccc}
& &  & & Intra-ring &  & & &  &  & Inter-ring &  & & \\
\end{tabular}
\begin{tabular}{cccc|ccc|cccc|ccc}
& $t_1$ & $t_2$ & $t_3$ \enskip \enskip \enskip  & $t^{'}_{1}$  &  \enskip  & $t^{'}_{2}$ \enskip \enskip \enskip \enskip  &  & $t_{1\perp}$ &
$t_{2\perp}$ & & $t^{'}_{1\perp}$ & $t^{'}_{2\perp}$ & \\ \hline
& -0.18 & -0.15 & -0.14 & -0.13 & & -0.14 & & -0.03 & -0.02 & & -0.02 & -0.03 &
\end{tabular}
\end{ruledtabular}
\end{table}
%===========================================================================

\begin{figure}
\includegraphics[width=9cm,keepaspectratio]{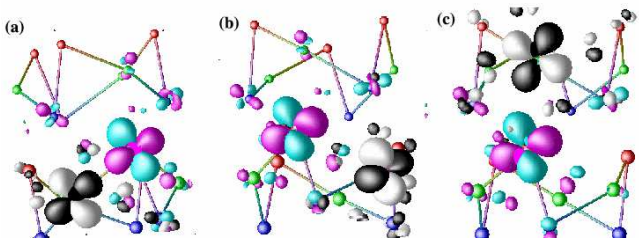}
\caption{ (Color on-line) Various intra-ring and inter-ring overlaps of the {\it downfolded}
V$-xy$ MTO-s at two V1 and V2 sites. 
%%Shown in the picture are two adjacent
%%V rings of the nanotube [corresponding to slices (A) and (B) in
%%FIG. 1].  
Plotted are the orbital shapes(constant-amplitude surfaces) with the lobes of opposite signs
  labeled by black(magenta) and white(cyan) at site V1(V2).
(a) Edge-sharing intra-ring, nearest-neighbor V1-V2 overlap ($t_1$) (b) Corner-sharing
intra-ring, second neighbor V1-V2 overlap ($t_1'$) (c) Inter-ring V1-V2 overlap ($t_{1 \perp}$).}
\label{downfolded}
\end{figure}

This result is very different from the extended H{\"u}ckel
molecular-orbital-based result of Whangbo and Koo\cite{Whangbo00}
which predicts that coupling via
corner-sharing pyramids is much larger than that via edge-sharing for
Na$_{2}$V$_{3}$O$_{7}$. While this conclusion is in general true for
vanadate systems like NaV$_{2}$O$_{5}$, LiV$_{2}$O$_{5}$ and
CsV$_{2}$O$_{5}$\cite{ueda} it fails in the case
of Na$_{2}$V$_{3}$O$_{7}$. Our result shows
that the peculiar geometry of the VO$_{5}$ coordination with
marked deviation of the V-O-V bond angles from 180$^{o}$
and highly non-coplanar nature of V-O-V bonds, gives rise to
coupling via edge-sharing nearly equally strong as that via
corner-sharing as long as one is confined to a single ring.  The  magnitude of intra-ring,
edge-sharing couplings are comparable to the largest edge-sharing
hopping
parameters in
other vanadate systems like
{CsV$_{2}$O$_{5}$ (-0.12 eV)}\cite{csvo} or  
{LiV$_{2}$O$_{5}$ (-0.18 eV)}\cite{livo}
% and
%\framebox{NaV$_{2}$O$_{5}$ (-0.18 eV)}\cite{navo}
while
the magnitude of intra-ring, corner-sharing couplings  is
much smaller than the largest corner-shared V-O-V in LiV$_{2}$O$_{5}$ or
NaV$_{2}$O$_{5}$ which are about -0.3  eV \cite{navo,livo}.
The corner-sharing coupling between two
adjacent rings is substantially diminished due
to further mis-alignment of V$-xy$ orbitals belonging to two different rings
which mixes the $\pi$
character of the bond with that of $\delta$ character.  
In FIG.\ \ref{downfolded} we show the various {\it downfolded} V$-xy$ MTO-s and their overlap
giving us an idea of the influence of the distorted geometry on the relative
orientations of the effective V$-xy$ orbitals, interaction paths and the
magnitude of overlaps.

%===========================================================================
%===========================================================================

Our  microscopically derived hopping integrals show that Na$_{2}$V$_{3}$O$_{7}$ can be described
as formed by tubes consisting of frustrated weakly coupled nine-site rings
with {\it partial} intra-ring frustration.
The {\it partial} frustration is
due to the absence of the second neighbor V1-V3 interaction in the ring which is neither
corner-sharing nor edge-sharing, rather it is decoupled by the intervening
V2(O$_{5}$) pyramid.

{\it Susceptibility -}  The exchange integral, J,  can be expressed in general as a sum of
antiferromagnetic and ferromagnetic contributions\cite{rosner} $J = J^{AF} + J^{FM}$. 
In the strongly correlated limit, typically valid for transition metal oxides, 
   the
 antiferromagnetic contributions are related to the hopping integrals,
$t_i$, by using a
perturbative approach as $J^{AF} = 4 t^2_i/(U-V_{inter})$, where $t_i$ corresponds to hopping
via various V-O-V superexchange paths and $U$ and $V_{inter}$ are the on-site and intersite
Coulomb interactions respectively. In the absence of a generally accepted satisfactory way
of direct computation of exchange integrals, such an approximate method
is
a good starting point
for estimates of the exchange couplings as well as
the relative strengths among the various exchange integrals\cite{footnote2}.

\begin{figure}[h]
\begin{center}
\includegraphics[width=6.5cm,angle=0]{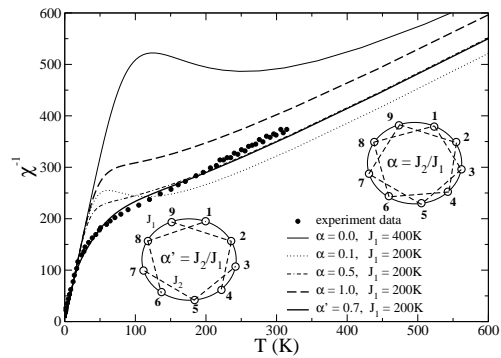}
\end{center}
\caption{Temperature dependence of the inverse
magnetic susceptibility in units of mol/emu obtained from exact
diagonalization for the two frustrated models of the inset
(see Eq.\ (\protect\ref{eq:susc2}))
compared with the experimental data\protect\cite{Gavilano03}
(filled dots).}
\label{fig:susc0}
\end{figure}

 In view of our {\it ab initio}
results,
the underlying spin-1/2 Hamiltonian for Na$_2$V$_3$O$_7$ can be written as
\begin{equation}
H \ =\  J_1\,\sum_{i=1}^9\,\left(
\vec{S}_i\cdot\vec{S}_{i+1} + \alpha_i\,\vec{S}_i\cdot\vec{S}_{i+2}
\right),\\
\label{eq:susc2}
\end{equation}

where we neglect the small differences in between the
three n.n. and two n.n.n intra-ring interactions and
to a first approximation also neglect the small inter-ring couplings.
$\alpha_i=J^i_2/J_1$ and we impose periodic boundary conditions
$\vec{S}_{L+i}=\vec{S}_i$ (L=9). In this model,
the n.n.n coupling is inhomogeneously
distributed in the sense that $\alpha_{i}$ = 0 for $i= $ 1, 4, 7 and
$\alpha_{i} \neq$ 0 (=$\alpha'$) for
$i= $ 2, 3, 5, 6, 8, 9 (see left illustration in
FIG.\ \ref{fig:susc0}).
For a further check of the goodness of this model we considered in addition
the fully frustrated model with $J^i_2=J_2$
for all $i$ ($\alpha_i=\alpha$)
(see right illustration in FIG.\ \ref{fig:susc0}).

%
% --------------------- % % --------------------- %
% --------------------- % % --------------------- %
%

In FIG.\ \ref{fig:susc0} we present a
comparison of the experimentally observed inverse magnetic
susceptibility\cite{Gavilano03} with that
obtained from exact diagonalization of the above-mentioned
two kinds of frustrated models, varying the parameters 
$J_1$ as well as $\alpha$ or $\alpha'$.

We observe that only the partially frustrated model ($\alpha$'$\neq$ 0),
which is the model predicted from the {\it ab initio} calculations
with $J_1$ and $J_2$ antiferromagnetically signed
is able to reproduce the experimental data over the whole range of
temperature. The fully frustrated model consistently shows, on the other
hand, an upturn of the inverse susceptibility at lower temperatures
not observed experimentally. We note here that estimates of $J_1$
and $J_2$ in terms of the downfolded $t_i$ values\cite{commentJ}
are within the same order of magnitude as those obtained here from the 
susceptibility comparison.

\begin{figure}[h]
\begin{center}
\vspace*{1cm}
\includegraphics[width=7cm,angle=0]{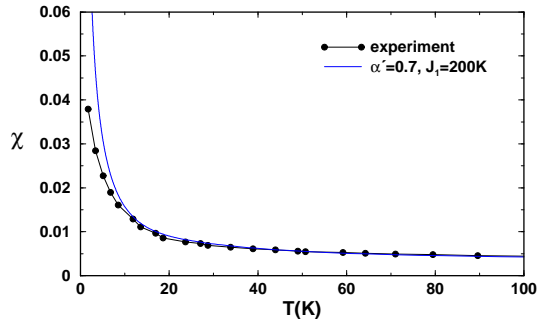}
%\includegraphics[width=6cm,angle=0]{susc2.eps}
%\hspace{0.5cm}\raise0.5cm\hbox{\includegraphics[width=1.1cm,angle=0]{model.eps}}
\end{center}
\vspace{-0.3cm}
\caption{
%Left panel:
%Inverse magnetic susceptibility in units of mol/emu obtained from exact
%diagonalization for the partially frustrated model for various values of
%$\alpha$' compared with  the experimental data\protect\cite{Gavilano03}
%(filled dots ).
Temperature dependence of the
susceptibility (emu/mol) at low temperatures.}
%%Right panel: Illustration of the
%%spin-diamond chain. Full and dashed lines correspond to $J_1$ and
%%$\alpha'J_1$ bonds respectively, see Eq.\ (\ref{eq:susc2}).}
\label{fig:susc2}
\end{figure}
% --------------------- % % --------------------- %

The partially frustrated model is invariant under
translations $i\to i+3$ ($i=1,\dots,9$) and corresponds
therefore in the low-energy sector to a three-site
ring with a  four-fold degenerate ground state, i.e.
it is a chirality-degenerate spin-doublet \cite{Kawano97}.
The susceptibility of the partially frustrated model
therefore diverges at low temperatures as
$1/T$ (illustrated in
FIG. \ \ref{fig:susc2}). Below $\sim 10\,{\rm K}$
the experimental susceptibility increases slower than
the calculated one suggesting
% expected
%for free spin-1/2 moments (one for every nine V-sites)
%indicating
the absence of a spin gap.
This is the range of energy where the inter-ring couplings, which have been neglected so far in our discussion,
will start to be important ( our {\it ab initio} $t_{\perp}=0.03eV$ corresponds
to a $J_{\perp}$ $\sim$ 10 K)

We consider in the following the contribution of the inter-ring couplings.
Our {\it ab initio} results show that -though  small-
there are various contributions, $t_{1 \perp}$,  $t_{2 \perp}$,
$t_{1 \perp}'$ and    $t_{2 \perp}'$,  to the inter-ring coupling
(see TABLE \ref{table} and FIG.\ \ref{hopping}) which will
be important at low temperatures
and can compete against each other.  These hoppings can be
taken as a reference for the important exchange integrals to
be considered, i.e. J$_{i \perp}$ and J$_{i \perp}'$. Assuming antiferromagnetic 
couplings, consideration of only  n.n.  J$_{\perp}$  
%The inter-ring exchange coupling $\sim 4 t_\perp^2/U$
will generally induce a spin-gap\cite{Kawano97} which will
scale with the inter-ring exchange coupling. On the other hand, the
existence
of two competing exchange integrals  J$_{\perp}$ and J$_{\perp}'$
which induce inter-ring
frustration
changes this scenario
completely.  As argued previously, the partially frustrated nine-site
ring model corresponds in the low-energy sector to a three-site ring.
We can therefore consider at low temperatures a model of three-site
rings
with frustrated inter-ring couplings which maps the main features
extracted from our {\it ab initio} calculations.  
The recent density matrix
renormalization group study  by
L{\"u}scher {\it et al.}\cite{Luscher04} on a tube of 3-site rings with
frustrated inter-ring couplings
showed that  the spin gap diminishes if
inter-ring frustration is present. For
perfect frustration, i.e. J$_{\perp}$=J$_{\perp}'$
the spin gap goes exactly to zero since then we are left with
an effective 1-dimensional S=1/2 Heisenberg model.
 Our {\it ab initio} hoppings $t_{\perp}$ and $t_{\perp}'$
are of the same strength what strongly points to
the condition of almost complete frustration
of the exchange integrals and therefore to a spin-gapless behavior
as has been observed
experimentally\cite{Gavilano03}.

Finally, it is worth mentioning here that the partially frustrated nine-site
ring model,
is of considerable interest by itself. For $\alpha'=1$ and large
number of sites it is geometrically
equivalent
to the spin-1/2 diamond chain, which corresponds in the low-energy
sector to a mixed spin-1 spin-1/2 chain \cite{mixed,pati}.
Our result $\alpha'=0.7$ indicates
that $\rm Na_2V_3O_7$ is close to the spin-diamond geometry.

%%In summary, by combining {\it ab initio} calculations with effective
%%model considerations, we are able to unveil the non-trivial electronic
%%and magnetic nature of the nanotubular system
%%Na$_{2}$V$_{3}$O$_{7}$.
%%Our {\it ab initio} calculation-based analysis shows  that frustration
%%is essential in this system.
%%We considered explicitly two regimes (i) moderate to high temperatures
%%where  a partially frustrated 9-site ring model
%%reproduces well the susceptibility behavior. (ii) low temperatures
%%where consideration of inter-ring couplings of frustrated nature
%%explains the absence of spin gap
%%in this system.

We acknowledge useful discussions with V.N. Muthukumar and J. Richter
and the support of the Deutsche Forschungs Gemeinschaft. We thank
the MPI-India partner group program for collaboration support.

%%%%%%%%%%%%%%%%%%%%%%%%%%%%%%%%%%%%%%%%%%%%%%%%%%%%%
%%%%%%%%%%%%%%%%%%%%%%%%%%%%%%%%%%%%%%%%%%%%%%%%%%%%%%

%%%%%%%%%%%%%%%%%%%%%%%%%%%%%%%%%%%%%%%%%%%%%%%%%%%%%%%%%%%%%%


\begin{thebibliography}{99}
\bibitem{footnote} We checked convergence of the band-structure
results obtained with the TB-LMTO scheme -where the downfolding
procedure is implemented- versus the results from the full
potential LAPW scheme.
\bibitem{note} The first neighbor shell of
intra-ring V-V pairs are of
distances: 2.91 ${\AA}$, 2.98 ${\AA}$ and 3.06 ${\AA}$.
The next set of intra-ring interactions are of distances
3.71 and 3.55 ${\AA}$.
\bibitem{footnote2}  To be precise, one should add to $J^{AF}$  the
$J^{FM}$ contribution as well, which 
may be estimated by evaluating the Coulomb 
exchange integrals between two Wannier functions centered at two 
neighboring sites. However, since we intend to extract from the {\it ab initio} 
calculation only the information of the important interaction pathes which 
should define the relevant exchange pathes, we have not included the FM 
contributions in our initial guess.
\bibitem{commentJ} For a U= 4 eV typical for vanadates and neglecting V$_{inter}$,
       we
obtain effective $J_1$= 285 K and $J_2$= 211 K. The ferromagnetic contribution
-not included here- will renormalize this estimate. We note that the
experimental $J_1$ is about 200 K.
\end{thebibliography}
\end{document}